\documentclass[iop]{emulateapj}

\usepackage{color}
\usepackage{graphicx}
\usepackage{amsmath}
\usepackage[colorlinks,linkcolor=black,anchorcolor=black,citecolor=black]{hyperref}

\newcommand\dx{{\rm d}}

\shorttitle{Dynamics of Inhomogeneous Dark Energy}
\shortauthors{Tian et al.}

\begin{document}
\title{The Dynamics of Inhomogeneous Dark Energy }
\author{ Shuxun Tian, Shuo Cao$^\ast$, and Zong-Hong Zhu}
\affil{Department of Astronomy, Beijing Normal University, 100875, Beijing, China; \emph{caoshuo@bnu.edu.cn}}

\begin{abstract}
In this paper, by analyzing the dynamics of the inhomogeneous
quintessence dark energy, we find that the gradient energy of dark
energy will oscillate and gradually vanish, which indicates the gradient energy of the scalar
field present in the early universe does not affect the current dynamics of the universe.
Moreover, with the decaying of gradient energy, there exists a
possible mutual transformation between kinetic energy and gradient
energy. In the framework of interacting dark energy models, we argue
that inhomogeneous dark energy may have a significant effect on the
evolution of the cosmic background, the investigation of which still
requires fully relativistic $N$-body numerical simulations in the
future.
\end{abstract}

\section{Introduction}\label{sec:01}

Since the observations of type Ia supernovae (SN Ia) first indicated
that the universe is undergoing an accelerated expansion at the
present stage \citep{Riess1998,Perlmutter1999}, dark energy, which
is generally believed to drive and fuel the cosmic acceleration, has
become one of the most important issues of the modern cosmology. In
the concordance standard cosmological model, dark energy acts as a
cosmological constant \citep[e.g.,][]{Amendola2010}, which was first
introduced by Einstein to obtain a static universe
\citep{Einstein1917}. Other candidate dark energy models include
quintessence \citep{Caldwell1998,Carroll1998,Zlatev1999}, phantom
\citep{Caldwell2002}, k-essence
\citep{Armendariz2000,Armendariz2001}, etc. Compared with the
cosmological constant model, the equation of state of quintessence,
a canonical scalar field $\phi$ with a potential $V(\phi)$, can
change over time. The tracker property found by
\citet{Zlatev1999} and \citet{Steinhardt1999} makes quintessence a good
candidate to alleviate the well-known coincidence problem of dark
energy, which is that the matter density is comparable with the dark energy
density today, whereas the matter density ($\rho_m$) decreases with
$a^{-3}$ and the cosmological constant density ($\rho_\Lambda$) does
not change in the cosmic expansion.

On the other hand, inflation can also be driven by a scalar field to
describe the accelerated expansion of the early universe
\citep{Liddle2000,Bassett2006,Linde2008,Linde2014,Baumann2015,Chernoff2015,Martin2016}.
Recently, increasing attention has been paid to the possibility of
whether a homogeneous and isotropic universe can be obtained through
inflation with relatively arbitrary initial conditions \citep[note that the initial conditions are mainly related to the initial value of the scalar field and its velocity; see][for a short
review]{Brandenberger2017}. In the framework of single scalar field
inflation models, it is generally believed that the cosmic expansion
can dilute the gradient energy and then generate an inflating
universe, following the homogeneous trajectory
\citep{Albrecht1987,Brandenberger1989,Kung1990,Muller1990,East2016}.
Things become much more interesting for the hybrid (multifield)
inflation models \citep[see][for a recent review]{Gong2017}.
\citet{Easther2014} found that subhorizon inhomogeneity could
enable some initial configurations, which cannot generate inflation
in the homogeneous limit, to realize inflation successfully.
However, this inhomogeneity could also deprive the inflation ability
of some other configurations, which could inflate in the homogeneous
limit.

Therefore, both inflation and the accelerated expansion of the late
universe can be driven by a scalar field. Similar to the case of
inflation, as the nature of dark energy is still unknown, whether it
is homogeneously distributed, and the size of the characteristic
length at which dark energy is homogeneously distributed, are still
pending issues. For instance, in the framework of perturbation
theories, it was found that dark energy with a Hubble scale
inhomogeneity can be used to explain the lower quadrupole moments in
the temperature fluctuation power spectrum of the cosmic microwave
background radiation \citep{Gordon2004,Gordon2005}. More recently,
\citet{Nunes2006} have studied the effect of dark energy with
cluster-scale inhomogeneity on the formation of large-scale
structures. However, it should be noted that in the previous works,
due to the absent backreaction mechanism, the first-order
perturbation term in the perturbation theory never affects the
evolution of the background. Therefore, it is necessary to study the
influence of the inhomogeneity of dark energy on the evolution of
the background universe, by using the method developed in the
inhomogeneous inflation scenarios. In addition to a different scalar field
potential to characterize dark energy, the effect of normal matter
still needs to be taken into consideration.

This paper is organized as follows. In Section \ref{sec:02}, we
briefly introduce the theory, the calculation method, and the
parameter setting in the calculation. In Section \ref{sec:03} we
present and discuss the main results. Our conclusions will be summarized
in Section \ref{sec:04}.

\section{Methodology}\label{sec:02}

In the form of Planck units, the action could be written as
\begin{equation}
  S=\int\dx^4x\sqrt{-g}\left[-\frac{M_{\rm P}^2}{2}R+\mathcal{L}_\phi+\mathcal{L}_m\right],
\end{equation}
where $M_{\rm P}=\sqrt{\hbar c/8\pi G}$ is the reduced Planck mass,
$\mathcal{L}_M$ is the Lagrangian for normal matter (including
radiation and dust), and
\begin{equation}
  \mathcal{L}_{\phi}=\frac{1}{2}g^{\mu\nu}\partial_\mu\phi\partial_\nu\phi+V(\phi)
\end{equation}
is the canonical Lagrangian density for the scalar field. Variation with
respect to the metric gives the Einstein field equations
\begin{equation}
  G_{\mu\nu}=\frac{1}{M_{\rm P}^2}\left[T^{(\phi)}_{\mu\nu}+T^{(m)}_{\mu\nu}\right],
\end{equation}
where
\begin{gather}
  T^{(\phi)}_{\mu\nu}=\partial_\mu\phi\partial_\nu\phi-g_{\mu\nu}\mathcal{L}_\phi,\\
  T^{(m)}_{\mu\nu}=\left(\rho+p\right)u_\mu u_\nu+pg_{\mu\nu}
\end{gather}
are the energy momentum tensor for the scalar field and matter,
respectively. The Klein-Gordon equation, which describes the motion
of the scalar field, could be expressed as
\begin{equation}
  \phi^{;\mu}_{\phantom{;\mu}{;\mu}}=\frac{\dx V}{\dx\phi},
\end{equation}
where the semicolon represents the covariance derivative.

\subsection{The Friedmann equation and Klein-Gordon equation}

In this paper, we will follow the method proposed in
\citet{Albrecht1987} and \citet{Easther2014} to study the dynamic properties of
the inhomogeneous dark energy. In this method, despite of the
inhomogeneous matter distribution, the universe could also be
described by a flat Friedmann-Lema\^{i}tre-Robertson-Walker (FLRW)
metric
\begin{equation}
  \dx s^2=-\dx t^2+a^2(\dx x^2+\dx y^2+\dx z^2),
\end{equation}
where $a=a(t)=a_0/(1+z)$ is the cosmological scale factor. This is
an accurate approximation when the inhomogeneity is small. In this
paper, we assume that normal matter is made up of dust and only
consider the one-dimensional inhomogeneous case for simplification,
i.e. $\phi=\phi(x,t)$. The extension of our analysis to the
three-dimensional case still calls for improved computational
capabilities in the future. Now the Friedmann equation is expressed as
\begin{equation}\label{eq:08}
  H^2=\frac{1}{3M_P^2}(\rho_\phi+\rho_m),
\end{equation}
where $H\equiv \dot{a}/a$ is Hubble parameter and $\dot{
}\equiv\frac{\partial}{\partial t}$. $\rho_\phi=\rho_\phi(t)$ is the
spatial average of the scalar field energy density
\begin{equation}
  \rho_\phi(t)=\frac{1}{L}\int_0^L\left[\frac{1}{2}\dot{\phi}^2+\frac{1}{2a^2}\phi'^2+V(\phi)\right]\dx x,
\end{equation}
where $'\equiv\frac{\partial}{\partial x}$ and $L$ is the spatial
scale selected in the numerical calculation. Note that in order to
eliminate the boundary effect, the value of $L$ should be much
larger than that of the characteristic scale of the inhomogeneity.
Combined with the Klein-Gordon equation, it is written as
\begin{equation}\label{eq:12}
  \ddot{\phi}+3H\dot{\phi}-\frac{1}{a^2}\phi''+\frac{\dx V}{\dx\phi}=0.
\end{equation}
Eq.~(\ref{eq:08}) and (\ref{eq:12}) could actually be used to
investigate the dynamics of the inhomogeneous dark energy. With the
definition of
\begin{equation}
  \Omega_m=\frac{\rho_m}{3H^2M_P^2},\quad\Omega_\phi=\frac{\rho_\phi}{3H^2M_P^2},
\end{equation}
the above equations could be used to describe the evolution of the
scalar field and the universe. The averaged gradient energy density
of the scalar field is defined as
\begin{equation}
  \rho_{\rm grad}(t)=\frac{1}{L}\int_0^L\frac{1}{2a^2}\phi'^2\dx x,
\end{equation}
which will be used throughout the following analysis in this paper.
We remark here that a reasonable approximation, i.e., the spatial
dependence of the metric, will not be taken into account in our
analysis, considering the small fraction of gradient energy in the
total background energy (which can be seen from Fig. \ref{fig:02}).
More specifically, the metric perturbation caused by the
inhomogeneous scalar field should be insignificant if the gradient
energy density is much smaller than the background homogeneous
energy density. Therefore, the influence of the metric perturbation
on the Klein-Gordon equation is negligible in our analysis.

\begin{figure*}
  \begin{center}
    \includegraphics[width=0.49\hsize]{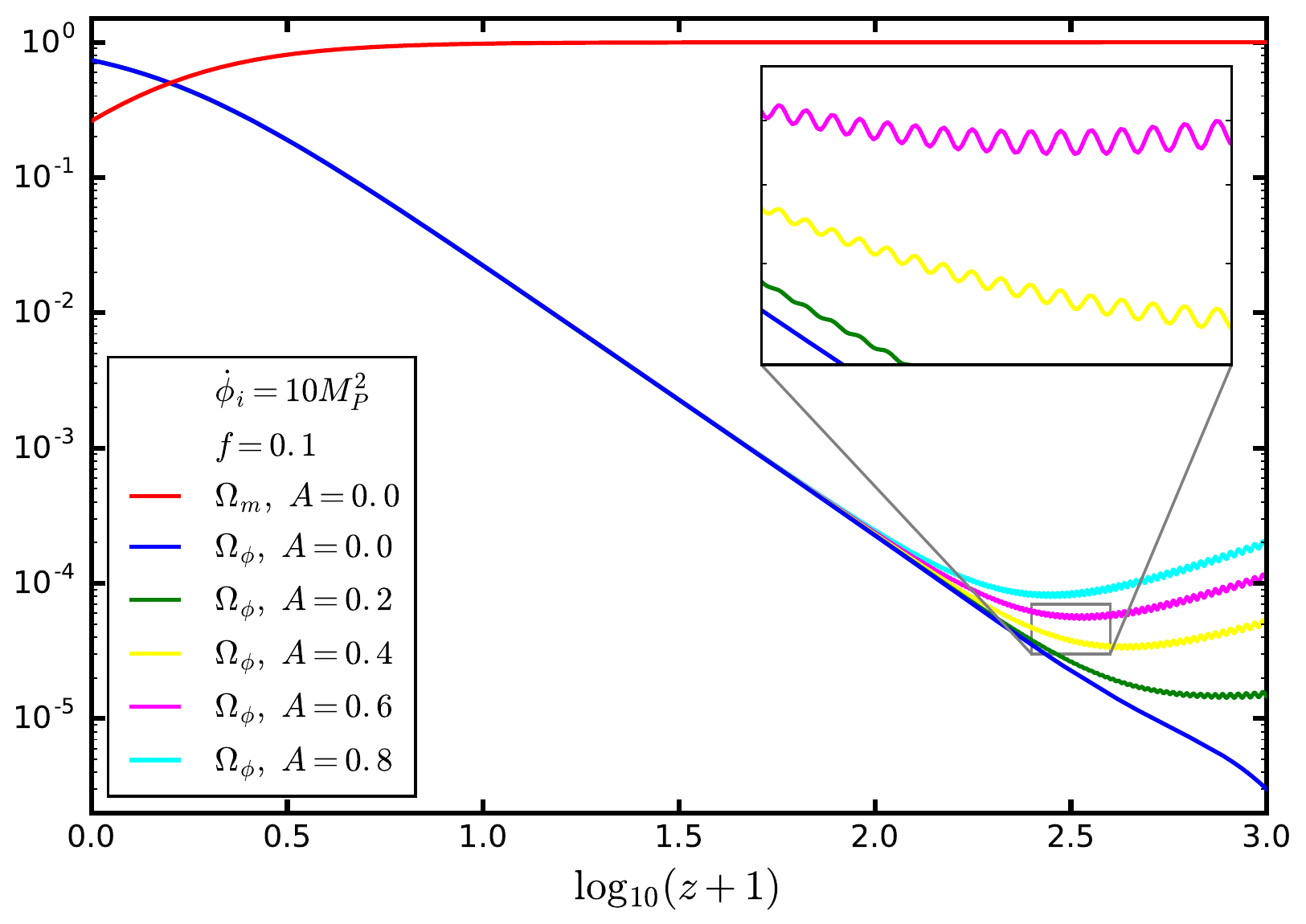}
    \includegraphics[width=0.49\hsize]{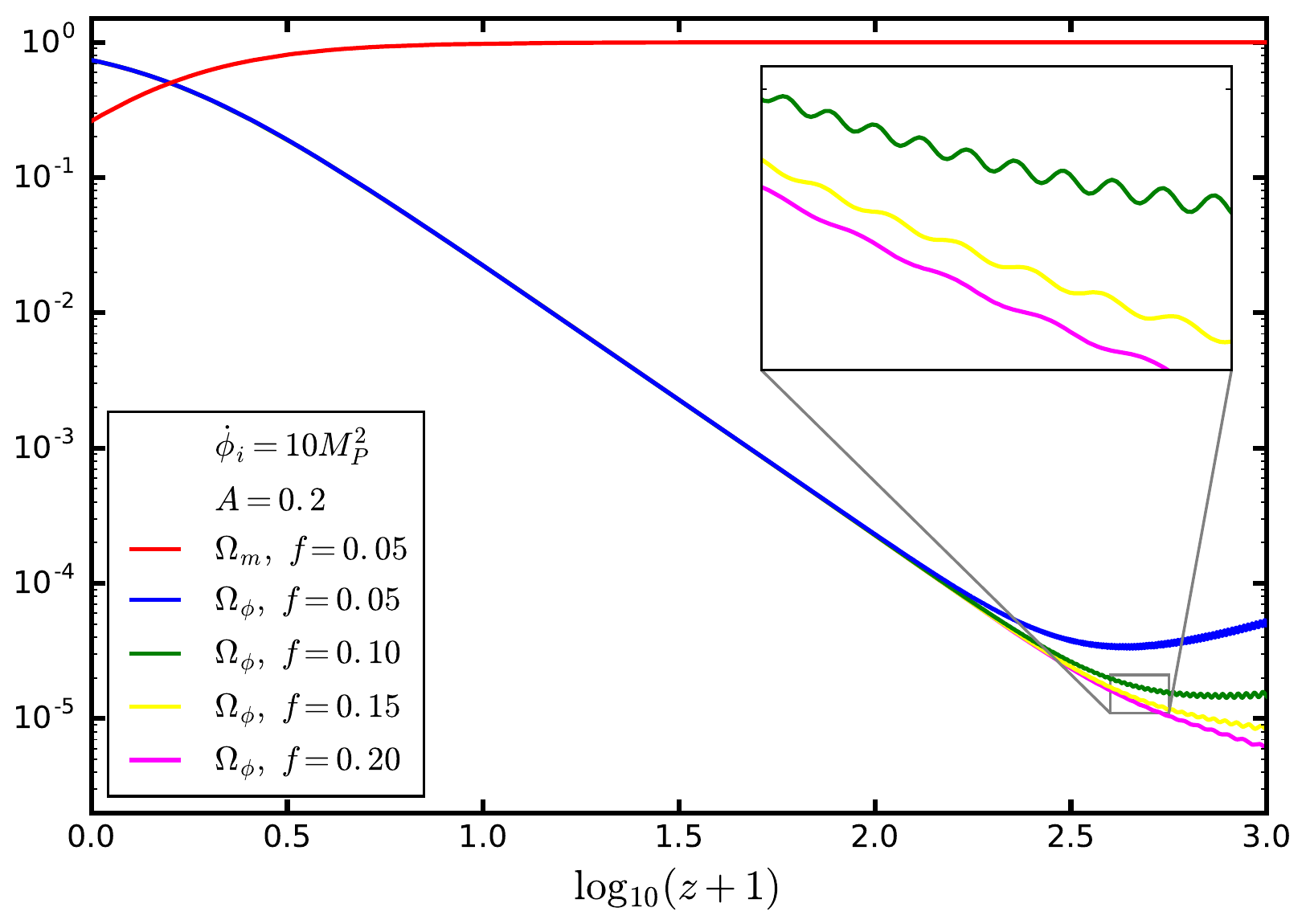}
  \end{center}
\caption{Evolution of $\Omega_m$ and $\Omega_\phi$ for different
combinations of $\{A,f\}$ with $\dot{\phi}_i=10M_P^2$. Details of the
evolution at early times are also shown in the insets. Left
panel: $f$ is fixed at 0.1, while $A$ varies from $0.0$ to $0.8$.
Right panel: $A$ is fixed at 0.2 while $f$ varies from $0.05$ to
$0.20$.}
  \label{fig:01}
\end{figure*}

\begin{figure}
  \centering
  \includegraphics[width=0.99\hsize]{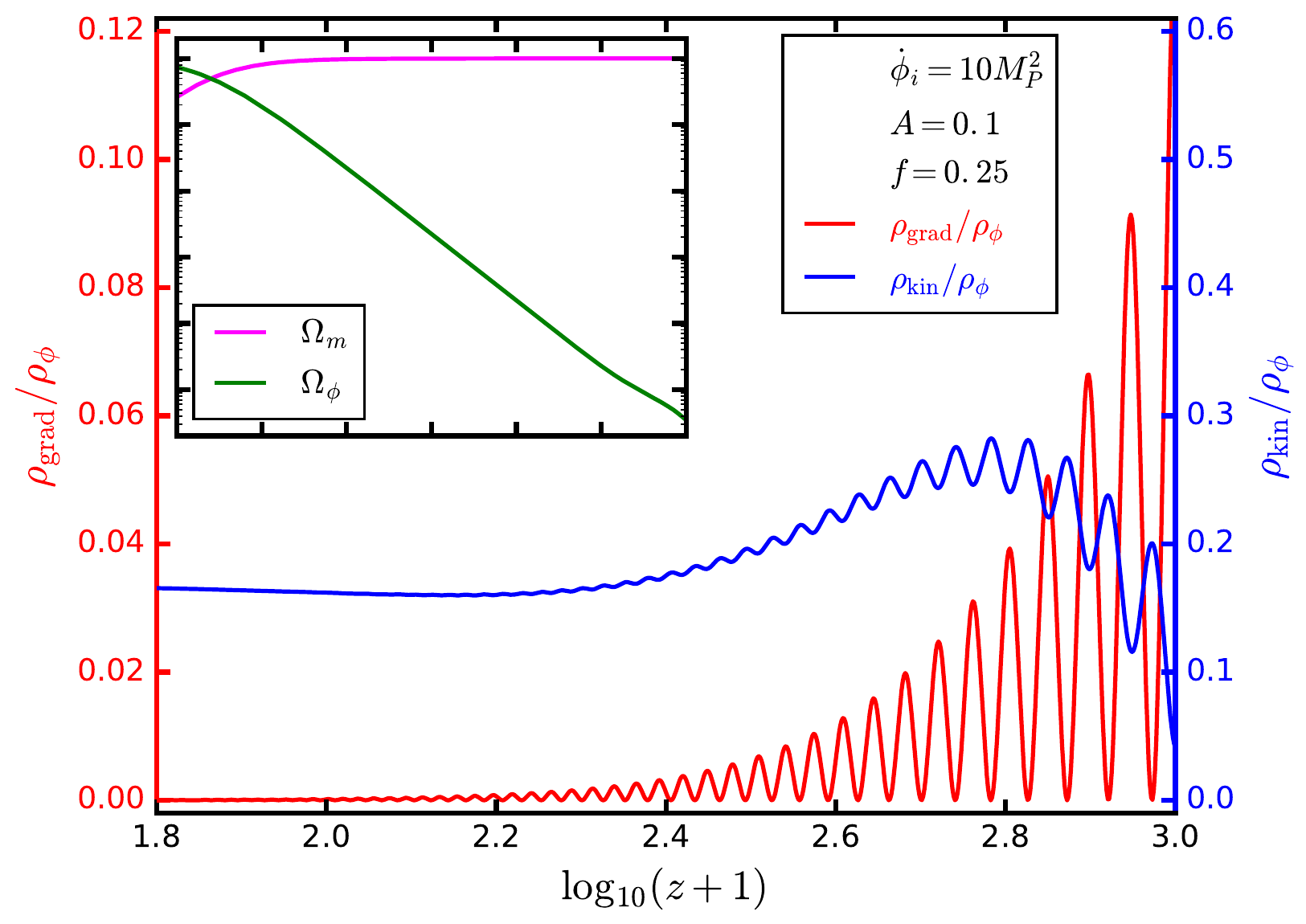}
\caption{Evolution of the ratio between the spatial averaged
gradient energy, and the total energy of the scalar field (red line,
$\rho_{\rm grad}/\rho_\phi$), the spatial averaged kinetic energy,
and the total energy of the scalar field (blue line, $\rho_{\rm
kin}/\rho_\phi$), with $\dot{\phi}_i=10M_P^2$, $A=0.1$, and
$f=0.25$. The inset shows the evolution of the variables
$\Omega_m$ and $\Omega_\phi$ for the same configurations. The range
of the $x$-axis in the inset is the same as that in
Fig.~\ref{fig:01}.}
  \label{fig:02}
\end{figure}

\subsection{Initial Conditions and Parameters}\label{sec:2.2}

We assume the current scale factor $a_0=1$. Compared with the
typical cosmological parameters derived from various observational
constraints ($H_0=70 {\rm km}/{\rm s}/{\rm Mpc}$,
$\Omega_{m0}=0.30$), we can set $\rho_{m0}=3.5\times10^{-121}M_P^4$
for $\rho_m=\rho_{m0}/a^3$, with cosmic evolution starting from
$a=10^{-3}$ and stopping at $a=1$. In this paper, we choose a simple
quintessence model with the potential
\begin{equation}
  V(\phi)=M^5\cdot\phi^{-1},
\end{equation}
where $M$ is a constant fixed at $M=10^{-24}M_P$ in order to be
closer to the real universe \citep[see, e.g.,][]{Amendola2010}. Based
on the assumption that there is only one mode of excited
inhomogeneity, the initial value of the inhomogeneous $\phi$ field
can be parameterized as
\begin{equation}
  \phi(x,t_i)=\phi_i[1+A\sin(2\pi kx)]
\end{equation}
where $\phi_i,A,k$ are non-negative constants. In our analysis, the
parameter $\phi_i$ is set at $\phi_i=10^{-3}M_P$ and two kinds of
initial velocity of the field are considered, $\dot{\phi}_i=10M_P^2$
and $10^3M_P^2$. Our approximate method requires $A<1$, which is
also the requirement of the quintessence model itself. Finally,
based on the assumption that the inhomogeneity is subhorizon, the
parameter $k$ could be written as $k=a_iH_i/f$, where $f<1$, $a_i$
is the initial scale factor, and $H_i$ is the initial Hubble
parameter. We choose $L=4/k$, which is large enough to eliminate the
boundary effect. One can verify that the parameters set in this work
will initially make the universe matter-dominated, which makes it
possible to estimate the initial Hubble parameter $H_i$ via the
energy density of the normal matter.

\subsection{Numerical Method}

As a system of integral-differential equations, Eq.~(\ref{eq:08})
and (\ref{eq:12}) need to be solved numerically. We firstly
discretize the $\phi$ field in the $x$ direction and then use the
discretized $\phi$ value to represent the differential and integral
with respect to $x$. Therefore, a series of coupled ordinary
differential equations will be obtained. Then we use the fourth-order
Runge-Kutta method to solve these ordinary differential
equations.

\begin{figure}
  \centering
  \includegraphics[width=0.99\hsize]{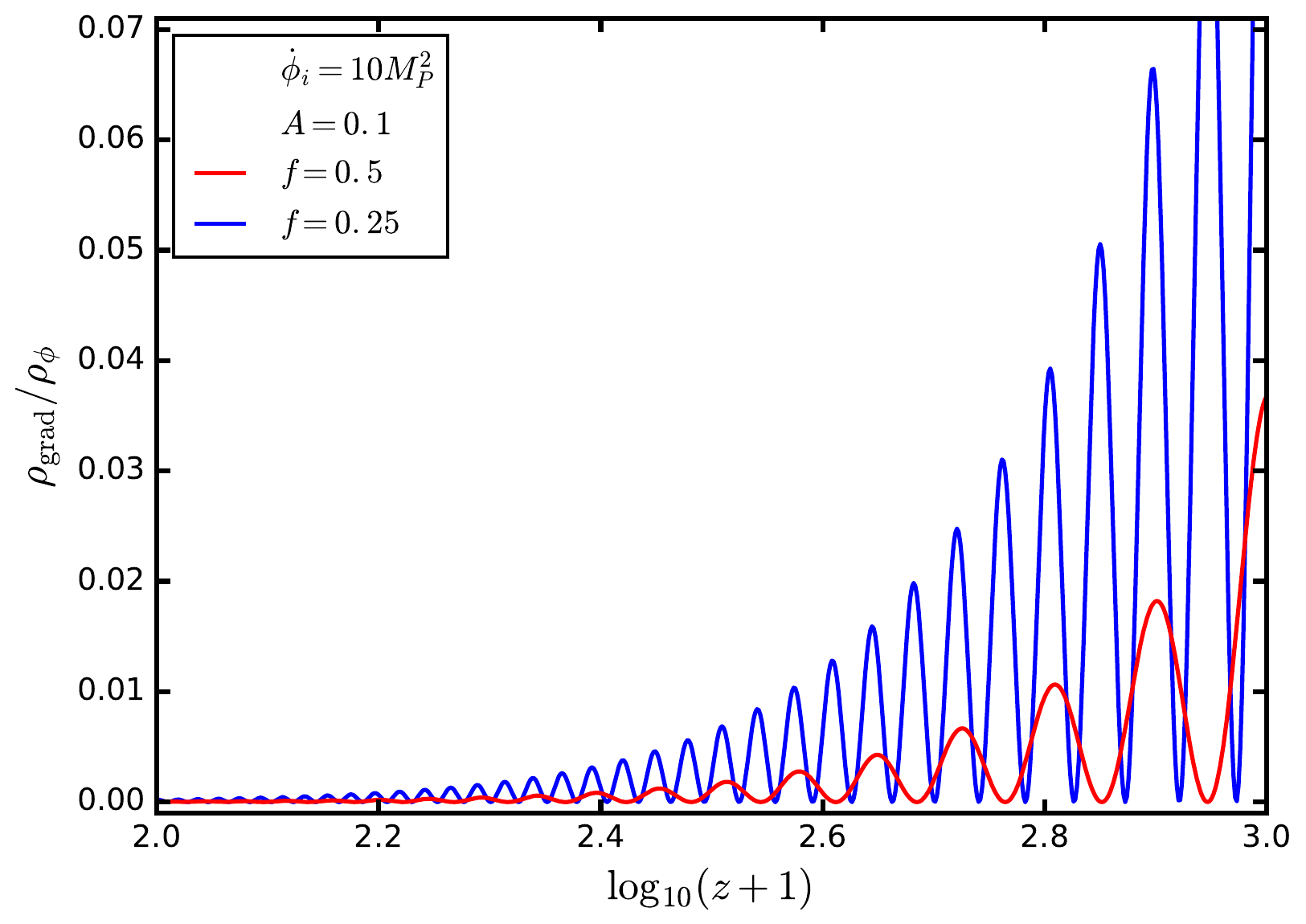}
\caption{Evolution of the ratio between the spatial averaged
gradient energy and the total energy of the scalar field with
$f=0.5$ (red line) and $f=0.25$ (blue line). The other two
parameters are fixed at $\dot{\phi}_i=10M_P^2$ and $A=0.1$.}
  \label{fig:03}
\end{figure}

\begin{figure*}
  \begin{center}
    \includegraphics[width=0.49\hsize]{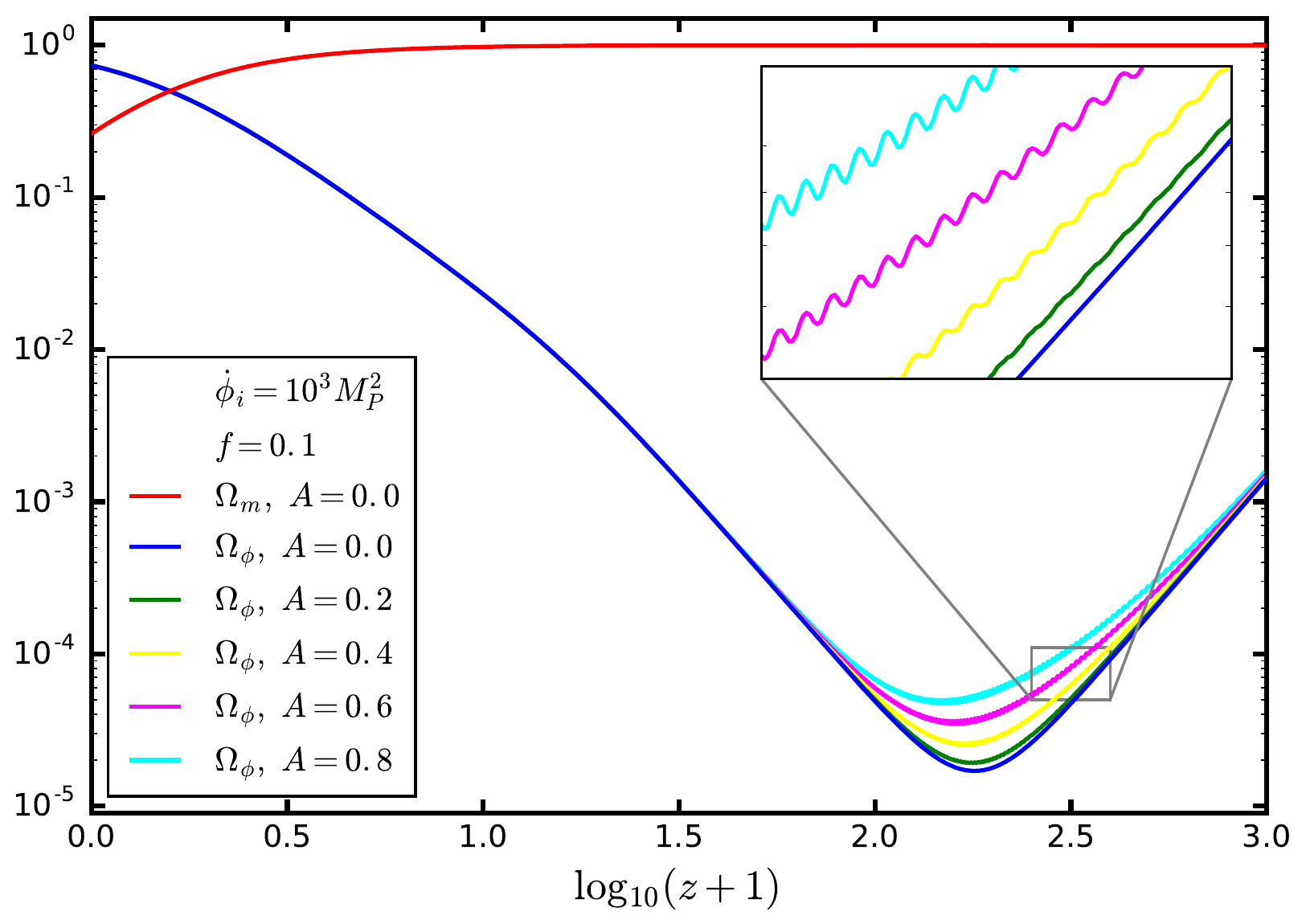} \includegraphics[width=0.49\hsize]{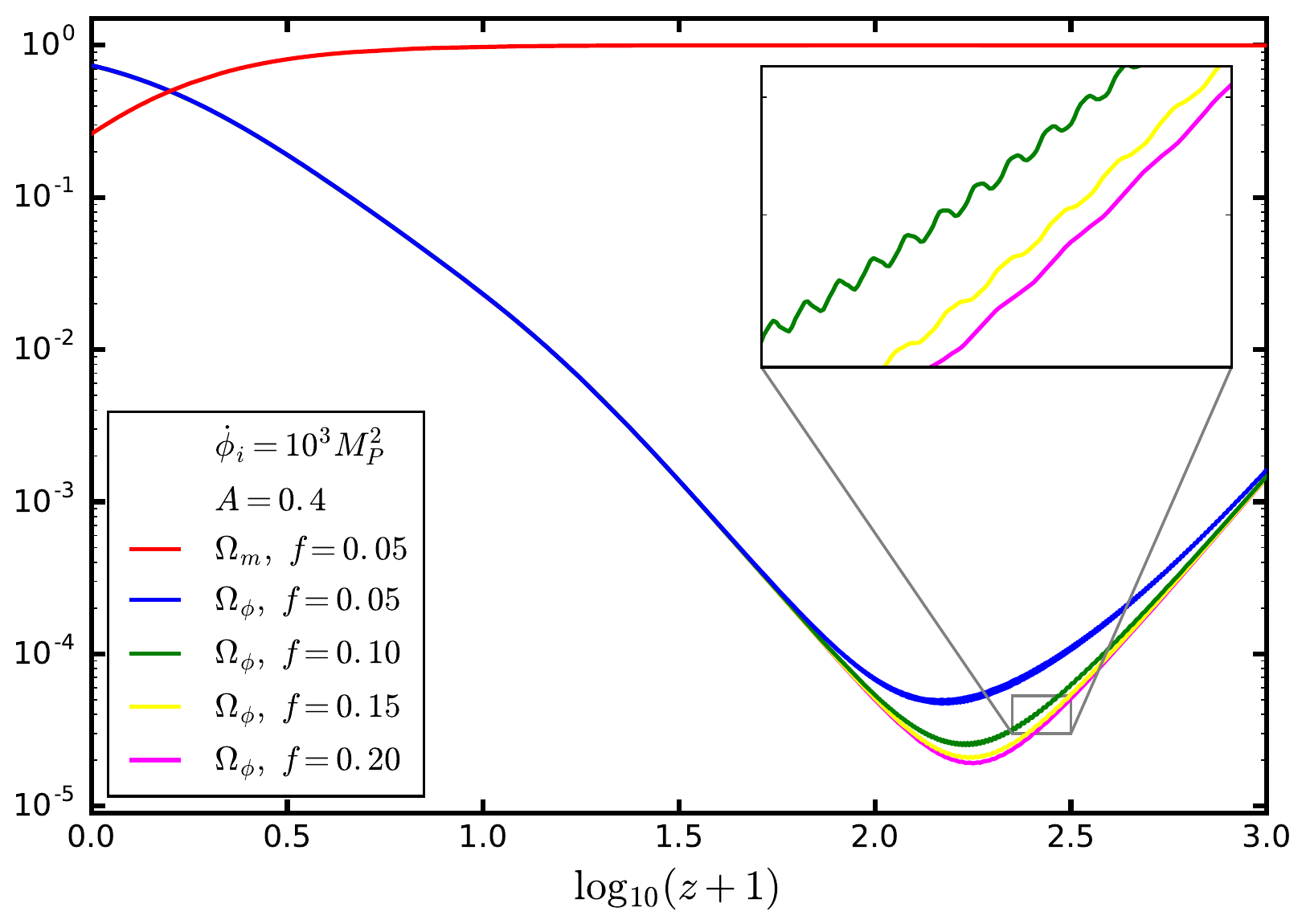}
  \end{center}
\caption{Same as Fig.~1 but with $\dot{\phi}_i=10^3M_P^2$.}
  \label{fig:04}
\end{figure*}

\section{Results and Discussions}\label{sec:03}

In order to study the effect of the gradient energy on the evolution
of the universe, we will carry out the calculation with different
combinations of  $\{A,f,\dot{\phi}_i\}$.

For the first case, the initial velocity is fixed at
$\dot{\phi}_i=10M_P^2$, which corresponds to the scalar field
dominated by potential energy initially in the homogeneous limit.
The evolution of $\Omega_m$ and $\Omega_\phi$ with respect to
redshift is plotted in Fig.~\ref{fig:01}, for different values of
$\{A,f\}$. It is obvious that the gradient energy vanishes with the
expansion of the universe, while the evolution of the universe
gradually tends to the homogeneous case ($A=0$). The presence of
gradient energy may also cause the scalar field density to
oscillate, which can be seen from the inset in
Fig.~\ref{fig:01}. More specifically, the frequency is hardly
affected by the changes of the oscillation amplitude (which
correspond to increasing value of $A$), while the decrease of $f$
may dramatically increase both the amplitude and frequency of the
oscillation.

In order to have a better illustration of the evolution of the
scalar field, in Fig.~\ref{fig:02} we show the ratio between the
mean kinetic energy density, the mean gradient energy, and the total
energy of the scalar field, with respect to redshift. The
corresponding scalar field parameters are set at $\{A=0.1,\ f=0.25,\
\dot{\phi}_i=10M_P^2\}$. We can clearly see the mutual
transformation between the kinetic energy and the gradient energy of
the scalar field, with the latter oscillating and decreasing with
time. In order to perceive the role played by the parameter $f$, we
also present the comparison results in Fig.~\ref{fig:03} with
$f=0.5, 0.25$, from which one could see that the decrease of $f$
will increase the oscillation frequency, a conclusion well
consistent with that obtained in Fig.~\ref{fig:01}.

For the second case, the energy density of quintessence at early
times increased in terms of its kinetic energy by
$\dot{\phi}_i=10^3M_P^2$, so that it does not need to be much less
than the matter density in the early universe, which provides an
attractive advantage compared with the cosmological constant
scenario. For this case we repeat the above calculation and obtain
the results shown in Fig.~\ref{fig:04}. Similarly, the gradient
energy vanishes with the expansion of the universe, while the
universe gradually tends to be homogeneous. This general property
should be considered in the cosmological models in which dark energy
has no directly interaction with normal matter, because the
expansion of the universe can actually dilute the gradient energy
and there is no mechanism to generate inhomogeneity of dark energy
in these models.

\section{Conclusions}\label{sec:04}

In this paper, we have studied the dynamics of the inhomogeneous
dark energy, in which dark energy is described by a canonical scalar
field. Our results show that with the expansion of the universe,
the gradient energy of the scalar field will oscillate and decrease
to zero, while the cosmic evolution gradually tends to follow the
homogeneous scenario. Therefore, the gradient energy of the scalar
field present in the early universe does not affect the current
dynamics of the universe. Although this conclusion is derived from
the analysis of a quintessence model, these general results could
also be applied to other cosmological models without the interaction
between dark energy and matter, including phantom, k-essence, etc.

Situations may change for the interacting dark energy models, which
take into account a possible interaction between dark energy and
dark matter through an interaction term
\citep{Bolotin2015,Wang2016}. Such cosmological scenarios have been
studied by many authors with different available observations
\citep{Cao11,Cao13,Salvatelli2014,Murgia2016,Nunes2016}. It is well
known that the distribution of matter is inhomogeneous on small
scales, and this inhomogeneity could be passed to the scalar field in
the interacting dark energy models. Despite of its low amplitude,
such small-scale (compared with the Hubble scale) inhomogeneity can
produce a large gradient energy and does not vanish with the
expansion of the universe, which is very likely to have a
significant effect on the dynamics of the cosmic background.
Recently, through $N$-body numerical simulations, the effect of the
inhomogeneity of the matter distribution on the evolution of a
matter-dominated universe or the standard $\Lambda$CDM universe has
been extensively investigated in many works
\citep{Adamek2016,Bentivegna2016,Giblin2016,Macpherson2017}. Their
results showed that the FLRW metric can well describe the evolution of
the background, i.e., the inhomogeneity of the matter distribution
has a negligible effect on the evolution of the cosmic background.
However, this conclusion is still controversial when considering the
backreaction theory, which has been extensively discussed in many
works  \citep{Buchert2012,Buchert2015,Racz2017}. More importantly,
the above conclusions still need to be verified by the
non-Newtonian simulations widely used in the literature
\citep{Eingorn2016,Hahn2016,Brilenkov2017}. To summarize, in the
framework of interacting dark energy models, the inhomogeneity of
the matter distribution may significantly influence the
inhomogeneity of the dark energy (scalar field) and thus the
evolution of the cosmic background. Such a possibility still needs to
be tested by fully relativistic $N$-body numerical simulations.

\vspace{0.5cm}

This work was supported by the Ministry of Science and Technology
National Basic Science Program (Project 973) under Grants No.
2014CB845806, the Strategic Priority Research Program ``The
Emergence of Cosmological Structure" of the Chinese Academy of
Sciences (No. XDB09000000), the National Natural Science Foundation
of China under Grant Nos. 11503001, 11373014, and 11690023, the
Fundamental Research Funds for the Central Universities and
Scientific Research Foundation of Beijing Normal University, China
Postdoctoral Science Foundation under grant No. 2015T80052, and the
Opening Project of Key Laboratory of Computational Astrophysics,
National Astronomical Observatories, Chinese Academy of Sciences.

\end{document}